\documentclass[12pt]{article}
\usepackage{amsmath,color}   
\usepackage{amssymb}
\usepackage{latexsym} 

\usepackage{graphicx}
\usepackage{epsfig}
\usepackage{hyperref}
\usepackage{cite}

\usepackage{tikz} %bp
\usetikzlibrary{cd} %bp

\usepackage[arrow,curve,matrix,arc]{xy}

\def\sign{{\rm sign}}

\newcommand{\bZ}{\mathbb{Z}}

\newcommand{\cQ}{\mathcal{Q}}

\newcommand{\kk}{k}

\newcommand{\eps}{\epsilon}
\newcommand{\vareps}{\varepsilon}

\newcommand{\ha}{{\wh{A}}}

\newcommand{\VV}{{\cal V}}

\newcommand{\NN}{{\cal N}}

\newcommand{\mathematica}[4]{\vspace{0.35cm}\noindent\boxed{\begin{minipage}{#1\textwidth}\begin{tabular}{lp{13cm}}{\color{paper_blue}{\scriptsize{\tt In[#4]:}}\raisebox{-0.65pt}{{\scriptsize{\tt=}}}}&{\tt #2}\\{\color{paper_blue}{\scriptsize {\tt Out[#4]:}}\raisebox{-0.65pt}{{\scriptsize{\tt=}}}}&{\tt #3}\end{tabular}\end{minipage}}\vspace{0.35cm}}

\newcommand{\be}{\begin{equation}}
\newcommand{\ee}{\end{equation}}
\newcommand{\ben}{\begin{eqnarray}\displaystyle}
\newcommand{\een}{\end{eqnarray}}

\newcommand{\refb}[1]{(\ref{#1})}
\newcommand{\p}{\partial}
\newcommand{\sectiono}[1]{\section{#1}\setcounter{equation}{0}}

\definecolor{varcolor}{rgb}{0.1,0.55,0.25}
\definecolor{functioncolor}{rgb}{0.1,0.35,0.75}
\definecolor{paper_blue}{rgb}{0.3,0.2,0.75}
\definecolor{paper_red}{rgb}{0.65,0.1,0.15}
\definecolor{paper_green}{rgb}{0.05,0.35,0.125}
\definecolor{paper_grey}{gray}{0.375}
\definecolor{perm}{rgb}{0.1,0.45,0.85}
\definecolor{deemph}{rgb}{0.7,0.7,0.7}

\newcommand{\cN}{\mathcal{N}}

\newcommand{\cM}{\mathcal{M}}
\def\Tr{\,{\rm Tr}\, }

\newcommand{\ta}{\tilde\alpha}

\renewcommand{\ha}{\hat\alpha}
\newcommand{\ca}{\bar \alpha}%\newcommand{\ca}{\check \alpha}
\newcommand{\cc}{\tilde c}
\newcommand{\dx}{c}

\newcommand{\gref}{g_C}

\newcommand{\OmS}{\Omega_{\rm S}}

\newcommand{\QC}{Q_C}
\newcommand{\bQC}{{\bar Q}_{C}}

\newcommand{\wt}{\widetilde}

\begin{document}

\baselineskip 18pt

\begin{center}
{\Large \bf  The Coulomb Branch Formula for Quiver Moduli Spaces}

\end{center}

\vskip .6cm
\medskip

\vspace*{4.0ex}

\baselineskip=18pt

\centerline{\large \rm Jan Manschot$^{1}$, Boris Pioline$^{2,3,4}$, Ashoke Sen$^{5}$}

\vspace*{4.0ex}

\begin{center}

{ \it $^1$ {\it Institut Camille Jordan, Universit\'e Claude Bernard 
Lyon 1, \\ 43 boulevard du 11 novembre 1918, 69622 Villeurbanne cedex, France} \\
$^2$ {\it CERN PH-TH,
Case C01600, CERN, CH-1211 Geneva 23, Switzerland}\\
$^3$ {\it Sorbonne Universit\'es, UPMC Univ. Paris 06, UMR 7589, LPTHE,
F-75005, Paris, France}\\
$^4$ {\it CNRS, UMR 7589, LPTHE, F-75005, Paris, France}\\
$^5$ Harish-Chandra Research Institute,
Chhatnag Road, Jhusi, Allahabad 211019, India
}

\end{center}

\vspace*{1.0ex}
\centerline{\small email\,:\,{\tt 
jan.manschot@univ-lyon1.fr, boris.pioline@cern.ch,
sen@hri.res.in}}

\vspace*{5.0ex}

\centerline{\bf Abstract} \bigskip

In recent series of works, by translating properties of multi-centered supersymmetric black holes
into the language of quiver representations, we proposed a  formula 
that expresses the Hodge numbers of the moduli space of semi-stable representations of 
quivers  with generic superpotential 
in terms of a set of  
invariants associated to `single-centered' or 
`pure-Higgs' states. The distinguishing 
feature of these invariants is that they are  independent of the choice of stability condition.
Furthermore they are uniquely determined by the $\chi_{y}$-genus of the moduli space.  %jm
Here,  we provide a self-contained summary of the Coulomb branch formula, spelling out mathematical details but leaving out proofs and physical motivations. 
 
\vfill \eject

\tableofcontents

\baselineskip=17pt

\sectiono{Introduction} \label{s1}
 
A quiver is a collection of nodes $i=1,\dots ,K$ connected by arrows $i\longrightarrow j$, and decorated by positive integers $N_i$ and  real parameters $\zeta_i$.\footnote{This definition differs slightly from the common definition of a quiver in mathematical literature. There a ``quiver'' is  typically defined as an oriented graph. The collection of data we call a ``quiver'', is in the mathematical literature sometimes referred to as a ``quiver setting''.} In the physics literature,
quivers describe the matter content of supersymmetric gauge theories \cite{Douglas:1996sw,Denef:2002ru}. The nodes label $U(N_i)$ gauge groups, the arrows denote bifundamental matter, while the $\zeta_i$'s are known as Fayet-Iliopoulos (FI)
parameters. In the mathematics literature, quivers are fundamental objects in representation theory (see e.g. \cite{DerksenWeyman, Reineke:2008} for entry points). 
The collection of integers $(N_1,\cdots N_K)$ is called the dimension vector 
and the parameters $(\zeta_1,\cdots \zeta_{K})$ define a choice of stability condition \cite{King:1993}.

We restrict to 2-acyclic quivers, i.e. such that there are no arrows  running in
both directions between any two nodes -- in a generic situation
 such arrows can be removed in pairs without
affecting the relevant results.  Let $\gamma_{ij}$ be the number of arrows from the $i$-th node to the $j$-th node: 
$\gamma_{ij}$ is a  positive integer if the arrows are directed
from the $i$-th to the $j$-th node, and negative in the opposite situation.
 There is no loss of generality in assuming that 
 the FI parameters associated to any quiver satisfy
\be \label{enzeta}
\sum_{i=1}^K N_i \zeta_i = 0\, .
\ee
It is convenient to introduce an abstract $K$-dimensional vector space spanned by
basis elements $\gamma_1,\cdots \gamma_K$, and associate a vector 
$\gamma=\sum_i N_i \gamma_i$ to a quiver with dimension vector $N\equiv
(N_1,\cdots N_K)$.
Since $N_i$'s are integers, 
the vector $\gamma$ belongs to  a $K$-dimensional lattice $\Gamma$ 
spanned by basis vectors $\gamma_i$.
We also denote by $\Gamma^+$ the cone of lattice vectors
of the form $\sum_i n_i\gamma_i$ with $n_i\ge 0$;  all physical quivers are 
described by some vector $\gamma\in\Gamma^+$. Finally
we introduce a vector space $\VV$ consisting of elements of the form
$\sum_i u_i\gamma_i$ with $u_i\in {\mathbb R}$, and
a bilinear symplectic product (the Dirac-Schwinger-Zwanziger, or DSZ product)
on $\VV$ via
\be\label{esymp}
\left\langle \sum_i u_i\gamma_i, \, \, \sum_j v_j\gamma_j\right\rangle=\sum_{i,j}
\gamma_{ij}u_i v_j\, .
\ee
It will also be convenient to introduce the inner product between the FI parameters
$\zeta=(\zeta_1,\cdots \zeta_K)$ and a vector $u=\sum_i u_i \gamma_i\in\VV$:
\be \label{einner}
\zeta \cdot u = \sum_{i=1}^K \zeta_i u_i\, ,
\ee
so that eq.\refb{enzeta} can be expressed as $\zeta\cdot\gamma=0$.
In the following we shall denote
by $\cQ(\gamma;\zeta)$ a quiver 
with dimension vector $ N$ and FI parameters $ \zeta$ satisfying \refb{enzeta}.  

To each quiver $\cQ(\gamma;\zeta)$, one associates a  quiver moduli space $\cM(\gamma;\zeta)$ defined as follows.
One introduces   complex
variables $\phi_{\ell\kk, \alpha, ss'}$ for every pair of nodes $\ell,\kk$ for which 
$\gamma_{\ell\kk}>0$. Here $\alpha$ runs over $\gamma_{\ell\kk}$ values,
$s$ is an index labelling the
fundamental representation of $U(N_\ell)$ and $s'$ is an index representing
the anti-fundamental representation of $U(N_{\kk})$. The quiver moduli space $\cM(\gamma;\zeta)$ 
is defined to be the  space of solutions to  the  D-term and F-term constraints,
\ben \label{emodi1}
&& \sum_{\kk , s,t,s',\alpha\atop \gamma_{\ell\kk}>0} \phi_{\ell\kk, \alpha, ss'}^* \, T^{(\ell)a}_{st} \, 
\phi_{\ell\kk,\alpha,t s'} - \sum_{\kk ,s,t,s',\alpha\atop \gamma_{\kk\ell}>0} 
\phi_{\kk\ell, \alpha, s's}^* 
\, T^{(\ell)a*}_{st} \, 
\phi_{\kk\ell,\alpha,s't} = \zeta_\ell \, \Tr(T^{(\ell)a})\quad \forall \, \ell, \, a \, , \nonumber  \\
&& {\p W\over \p \phi_{\ell\kk,\alpha,ss'}}=0, \quad \forall \, \,  \ell, k, \alpha, s, s'\, \ ,
\een
modded out by the natural action of the gauge group $\prod_\ell U(N_\ell)$. 
Here $T^{(\ell)a}$'s are the generators of the $U(N_\ell)$ gauge group, and $W$ is a
gauge invariant holomorphic function of the variables
$ \phi_{\ell\kk, \alpha, ss'}$, known as the superpotential\footnote{We consider only 
generic superpotentials, which contain a generic linear combination
of all the generators of the ring of gauge invariant polynomials of the variables
$\phi_{\ell k,\alpha, ss'}$. The Hodge numbers of the quiver moduli space are then independent
of the choice of superpotential, and we omit the dependence of $\cM(\gamma;\zeta)$ on $W$.}.%bp
Equivalently, one may define $\cM(\gamma;\zeta)$
as the quotient of a certain open subset of solutions of the F-term constraints
(the second line in \eqref{emodi1}) by the action of the complexified gauge group
$\prod_\ell GL(N_\ell,\mathbb{C})$. The open subset is the space of 
quiver representations which are semi-stable with respect to $\zeta$\cite{King:1993}.
This induces a natural complex structure on $\cM(\gamma;\zeta)$, and shows that
$\cM(\gamma;\zeta)$ is a quasi-projective algebraic variety. For generic 
superpotential and primitive charge vector $\gamma$ (i.e. ${\rm gcd}(\{N_i\})=1$), 
$\cM(\gamma;\zeta)$ is believed to be a compact, smooth projective variety.

Let $h^{p,q}(\cM)$ denote the Hodge numbers of the quiver moduli space $\cM(\gamma;\zeta)$. We define the  Dolbeault polynomial of $\cM(\gamma;\zeta)$ to be
\be \label{eDol}
Q(\gamma; \zeta; y,t) \equiv \sum_{p,q} h^{p,q}(\cM) \, (-y)^{p+q-d} \, t^{p-q}\, ,
\ee 
where $y$ and $t$ are complex variables. Thus $Q(\gamma; \zeta; y,t)$ encodes all the Hodge numbers of $\cM(\gamma;\zeta)$. %bp
The specialization 
of $Q(\gamma; \zeta; y,t)$ at $y=1/t=v^{1/2}$ is proportional to the Hirzebruch $\chi_y$-genus $\chi(\cM, y)=\sum_{p,q}h^{p,q}(\cM) \, (-1)^{p} \, y^{q}$, %jm
\be \label{eHir}
Q(\gamma; \zeta; y=v^{1/2};t=v^{-1/2}) \equiv \sum_{p,q} h^{p,q}(\cM) \, (-1)^{p+q-d} \, v^{q - d/2}
= (-1)^d \, v^{-d/2} \chi(\cM, -v)
\ee
The Hirzebruch $\chi_y$-genus can in principle be computed using the Hirzebruch-Riemann-Roch theorem (see {\it e.g.} \cite{1310.1265}). A systematic procedure for computing it using localization techniques in quiver quantum mechanics was  presented recently in \cite{Cordova:2014oxa,Hori:2014tda}. %jm

 In order to deal with cases where 
$\gamma$ is not primitive, it is useful to define 
\be \label{edefqbar}
\bar Q(\gamma; \zeta;y,t) = \sum_{m|\gamma} 
\frac{1}{ m}  {y - y^{-1}\over y^m - y^{-m}}
Q(\gamma/m; \zeta;y^m, t^m)\, ,
\ee
where $m|\gamma$ means that $m$ is a common divisor of
$(N_1,\cdots , N_K)$ if $\gamma =\sum_\ell N_\ell \gamma_\ell$.
$\bar Q(\gamma; \zeta;y,t)$ is a polynomial in $t$ with coefficients which are rational functions of $y$.
The Dolbeault polynomial $Q(\gamma; \zeta;y,t)$ can be recovered from $\bar Q(\gamma; \zeta;y,t)$ by the M\"obius inversion formula,
\be
\label{QtobQ}
Q(\gamma; \zeta;y,t) = \sum_{m|\gamma} 
\frac{\mu(m)}{ m}  {y - y^{-1}\over y^m - y^{-m}}
\bar Q(\gamma/m; \zeta;y^m, t^m)\, ,
\ee
 where $\mu(m)$ is the M\"obius function (i.e. 1 if $m$ is a product of an even number of 
distinct primes, $-1$ if $m$ is a product of an odd number of distinct
primes, or $0$ otherwise). For primitive $\gamma$, $Q$ and $\bar Q$ coincide. 
If $\gamma$ is not primitive, the moduli space $\cM(\gamma;\zeta)$ 
is not expected to be compact and the algebraic-geometric definition of either 
$Q$ or $\bar Q$ is more involved \cite{Joyce:2004,
  Kontsevich:2008}. The coefficients of $Q$ are expected to give the intersection cohomology of the compactified
moduli space. Moreover, $\bar Q$ is a particularly useful construct as it
appears in the wall-crossing formula \refb{ewall} even for quivers with primitive total charge. 
 
The Coulomb branch formula proposed in 
\cite{1103.1887,1207.2230,1302.5498} is a conjectural 
formula, valid for an arbitrary quiver, which expresses 
$Q(\gamma; \zeta; y,t)$  in terms of a set of invariants $\OmS(\alpha;t)$, %bp
which are independent of the stability condition $\zeta$ and (for a generic superpotential $W$)
of the complex parameter $y$. We refer to the $\OmS(\alpha;t)$ as the %bp
`single-centered' or `pure-Higgs' invariants. An independent mathematical definition of these invariants would be highly desirable. %bp
This formula was motivated from the study of multi-centered black hole quantum mechanics
in $\NN=2$ supergravity. Our goal in this note is to explain this formula in mathematical detail,
leaving out physical motivations. For the physical reasoning that led to the formula, and for the proof of some of its properties, the reader is referred to the original papers. More pedagogical reviews
can be found in \cite{1304.7159,Diab}. As an aid to the reader, we represent below the dependency of the main ingredients which enter the Coulomb branch formula, with equation numbers where the relations can be found: %as below
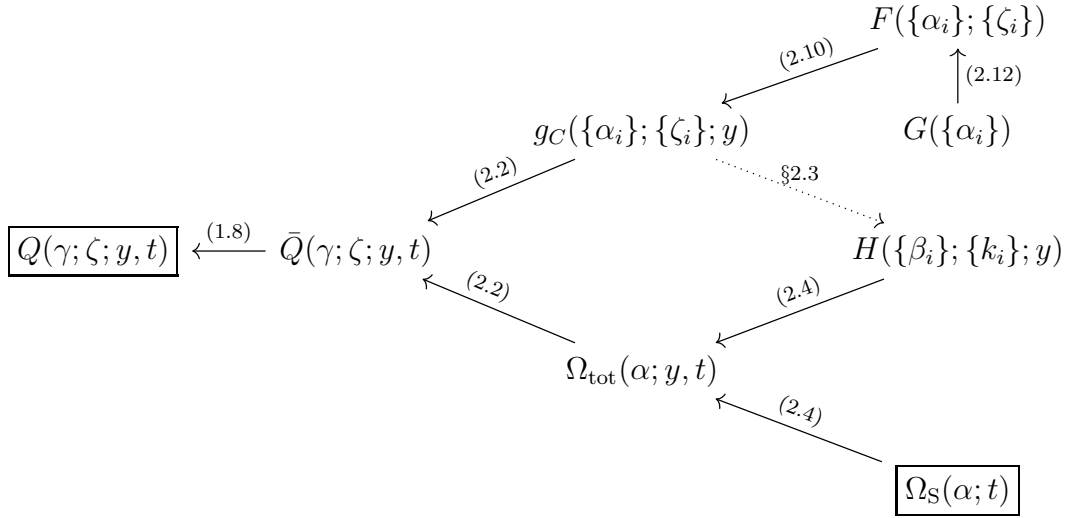
\begin{center}
\begin{tikzcd}
& & & F(\{\alpha_i\};\{\zeta_i\})  \arrow[leftarrow]{d}{\eqref{enewrecur}} \\
& & \gref(\{\alpha_i\};\{\zeta_i\};y) \arrow[leftarrow]{ru}[sloped,pos=0.8]{\eqref{eqforg}} & G(\{\alpha_i\}) \\
\boxed{Q(\gamma;\zeta;y,t)} \arrow[leftarrow]{r}{\eqref{QtobQ}} & 
{\bar Q}(\gamma;\zeta;y,t) \arrow[leftarrow]{ru}[sloped,pos=0.7]{\eqref{essp1}} 
                                             \arrow[leftarrow]{rd}[sloped,pos=0.2]{\eqref{essp1}}  & & 
         H(\{\beta_i\};\{k_i\};y) \arrow[lu,dotted,leftarrow,"\S 2.3" above] \\
 & & \Omega_{\rm tot}(\alpha;y,t) \arrow[leftarrow]{ru}[sloped,pos=0.7]{\eqref{essp2}} 
                                    \arrow[leftarrow]{rd}[sloped,pos=0.3]{\eqref{essp2}}  &  \\
 & & & \boxed{\OmS(\alpha;t)} \\
\end{tikzcd}
\end{center}

Finally a word about notation: throughout our discussion we shall reserve the symbols
$\gamma_1,\cdots \gamma_K$ for the basis vectors of $\Gamma^+$ as introduced above.
On the other hand the symbols $\gamma$, $\alpha_i$, $\beta_i$ etc. will be used to label
more general elements of $\Gamma^+$ and occasionally also elements of the vector
space $\VV$ which do not necessarily lie in $\Gamma^+$. 

\sectiono{Statement of the Coulomb branch formula} \label{s2}

The Coulomb branch formula expresses $Q(\gamma; \zeta; y,t)$ in terms of a set of 
`single-centered' or `pure-Higgs' indices $\OmS(\alpha;t)$.
Since this formula has not been proven in full generality, we shall denote it by $\QC(\gamma; \zeta; y,t)$
with the understanding that $\QC$ is conjecturally equal to $Q$.\footnote{The functions
$\QC$ and $\gref$ were denoted by $Q_{\rm Coulomb}$ and $g_{\rm Coulomb}$ in
\cite{1302.5498}.}
The formula will only be valid away from the walls
of marginal stability, \i.e.\ for choices of FI parameters $\zeta$ for which
\be 
\sum_{i=1}^K n_i \zeta_i \ne 0 \quad \hbox{for} \quad 0\le n_i \le N_i
\ee
except when $n_i$ is proportional to $N_i$. %bp
On a wall of marginal stability the moduli space $\cM$ and hence the associated
Dolbeault polynomial are ill-defined.

 \subsection{The main formula} \label{smain}

The proposed formula for $\bQC(\gamma; \zeta; y,t)$ takes the form
\ben
 \label{essp1}
%\QC(\gamma; \zeta;y,t) &=& \sum_{m|\gamma} 
%\frac{\mu(m)}{ m}  {y - y^{-1}\over y^m - y^{-m}}
%\bQC(\gamma/m; \zeta;y^m, t^m)\, ,
%\nonumber \\
\bQC(\gamma; \zeta;y,t) &=&
\sum_{n\ge 1}\sum_{\{\alpha_i\in \Gamma^+\} \atop \sum_{i=1}^n \alpha_i =\gamma}
{1\over n!} \, \gref\left(\{\alpha_1, \cdots, \alpha_n\},
 \{\dx_1,\cdots \dx_n\};y\right)
\nonumber \\ && \quad
\prod_{i=1}^n \left\{\sum_{m_i\in\bZ\atop m_i|\alpha_i}
{1\over m_i} {y - y^{-1}\over y^{m_i} - y^{-m_i}}\, 
\Omega_{\rm tot}(\alpha_i/m_i;y^{m_i}, t^{m_i})
\right\}
\, ,
\een
where the sums over $n$ and $\{\alpha_1,\cdots \alpha_n\}$  label all possible ways of
expressing $\gamma$ as {\it ordered} sums of elements $\alpha_i$ of $\Gamma^+$. 
The functions $\gref(\{\alpha_1,\cdots, \alpha_n\};
 \{\dx_1,\cdots \dx_n\};y)$, known as Coulomb indices, will be given explicitly 
in \S\ref{sdefg}.\footnote{Since this function is symmetric in the arguments $\{\alpha_1,\cdots \alpha_n\}$,
 we can replace the sum over the ordered set $\{\alpha_1,\cdots, \alpha_n\}$ by
 the sum over an unordered set, and replace the $n!$ in the denominator by  
a symmetry factor  $|{\rm Aut}(\{\alpha_1,\cdots, \alpha_n\})|$  given by
$\prod_k s_k!$ if among the set $\{\alpha_i\}$ there are
$s_1$ identical vectors $\tilde \alpha_1$, $s_2$ identical vectors
$\tilde\alpha_2$ etc.}
The $c_i$'s in the argument of $g_C$ are given in terms of the FI parameters by
\be
c_i = \zeta\cdot\alpha_i
\, .
\ee
The condition \refb{enzeta} together with $\sum_i\alpha_i=\gamma$ guarantees that
$\sum_i c_i=0$.
Finally the functions $\Omega_{\rm tot}(\alpha;y,t)$ 
are given by
\be \label{essp2}
\Omega_{\rm tot}(\alpha;y,t) = \OmS(\alpha;t) + 
\sum_{\{\beta_i\in \Gamma^+\}, \{m_i\in\bZ\}\atop
m_i\ge 1, \, \sum_i m_i\beta_i =\alpha}
H(\{\beta_i\}; \{m_i\};y) \, \prod_i 
\OmS(\beta_i;  t^{m_i})
\, .
\ee
The sum over $\{\beta_i\}$ and $\{m_i\}$ 
run over {\it unordered} sets satisfying the constraint $\sum_i m_i\beta_i =\alpha$.
$H(\{\beta_i\}; \{m_i\};y)$ are functions which can be determined from the functions $\gref$ following
an algorithm to be described in \S\ref{sdefh}. The functions $\OmS(\alpha;t)$, called single centered
indices, are unknown functions
of $t$ but are independent of $y$ and $\zeta$.
They are required to satisfy $\OmS(\alpha;t)=0$ for $\alpha\not\in\Gamma^+$.
They also have no explicit dependence on the vector $\gamma$ labelling the original quiver;
so for different $\gamma$'s we use the same set of $\OmS(\alpha;t)$'s.
Some further constraints on $\OmS(\alpha;t)$, as well as their determination from the
Hirzebruch $\chi_y$-genus, will be discussed in \S\ref{soms}. %jm

\subsection{The Coulomb index $\gref$} \label{sdefg}

Our goal in this subsection is to give an expression for $\gref\left(\{\alpha_1, \cdots, \alpha_n\},
 \{\dx_1,\cdots \dx_n\};y\right)$. 
 First we define
 \be 
 \alpha_{ij} =\langle \alpha_i, \alpha_j\rangle\, .
 \ee
 There are two equivalent definitions of $\gref$. The first one is simple to state but involves
testing the existence of solutions of 
some non-linear algebraic equations. The second one is an iterative procedure that
is more complicated to state but is purely combinatorical. %bp
In both definitions we have, for a single charge vector,
\be
g_C(\{\alpha_1\}; \{c_1=0\}; y)=1 \quad \forall \, \alpha_1\, .
\ee

\subsubsection{First definition} \label{sfirst}
 
%$\gref(\{\alpha_1,\cdots \alpha_n\}; \{\dx_1,\cdots \dx_n\};y)$  is given by
 \be \label{eindex1}
\gref(\{\alpha_1,\cdots \alpha_n\}; \{\dx_1,\cdots \dx_n\};y) 
= (-1)^{n-1+\sum_{i<j} \alpha_{ij} } (y-y^{-1})^{-n+1} \sum_{{\rm solutions} \, P}\, s(P) 
\, y^{\sum_{i<j} 
\alpha_{ij} \, {\rm sign}(z_j-z_i)} \, ,
\ee
where the sum runs over solutions to the equations
\be \label{eor1}
\sum_{j=1\atop j\ne i}^n {\alpha_{ij}\over |z_i - z_j|}  =  \dx_i\, , \quad \hbox{for $1\le i\le n-1$}\, \ ,
\quad z_1=0\ .
\ee
$s(P)$ in \eqref{eindex1} is given by the sign of the Hessian 
$\det (\p^2 V / \p z_i \p z_j)$  
of the function
\be \label{ecou0}
V(\{z_i\})=-\sum_{i,j=1\atop i<j}^n \alpha_{ij} \, {\rm sign}(z_j-z_i) \log| z_i- z_j|
-\sum_{i=1}^n \dx_i\, z_i \, ,
\ee
whose critical points reproduce the conditions \eqref{eor1}.  
This prescription follows from computing the index of the Dirac operator 
on the moduli space of multi-centered black holes
using the Atiyah-Bott-Lefschetz formula \cite{1011.1258,1103.1887}.

Eq.\refb{eindex1} gives an unambiguous definition of 
$\gref(\{\alpha_1,\cdots \alpha_n\}; \{\dx_1,\cdots \dx_n\};y)$ when each solution to 
\refb{eor1} has all the centers distinct. However the enumeration of distinct solutions becomes
ambiguous when two or more centers coincide.\footnote{Since $1/|z_i-z_j|$ diverges as
$z_i\to z_j$, to examine the existence of solutions with coincident centers we need to
take $|z_i-z_j|\sim \eps$ for a subset of the centers and then examine if it is possible to
satisfy \refb{eor1} in the $\eps\to 0$ limit for suitable choice of ratios of these distances. 
These were called collinear scaling solutions
in \cite{1302.5498}.}
The remedy found in \cite{1302.5498}
is to work with a deformed set of parameters. The rules for generating these deformations 
will be reviewed in \S\ref{sdeform}.

The algorithm described above
requires us to find all solutions to \refb{eor1} and hence in general 
can only be
carried out numerically. (Note however that the final expression \refb{eindex1} is 
insensitive to the details
of the solution and depends only on the relative ordering of the $z_i$'s in a given solution).
There is however an equivalent definition of 
$\gref(\{\alpha_1,\cdots \alpha_n\}; \{\dx_1,\cdots \dx_n\};y)$, derived from 
the above definition, in which each step 
can be carried out analytically.  We shall now give this alternative definition, but casual readers
can jump to \S\ref{sdefh} for the definition %bp
of the function $H$ appearing in \refb{essp2}.

\subsubsection{Second definition} \label{second}

The alternate definition of $\gref$ takes the form
\ben \label{eqforg} 
&& \gref(\{\alpha_1,\cdots \alpha_n\}; \{\dx_1,\cdots \dx_n\};
y) \nonumber \\
&=& (-1)^{n-1+\sum_{i<j}\alpha_{ij}}
(y-y^{-1})^{-n+1}  \sum_{\sigma}
F\left(\{\alpha_{\sigma(1)},\cdots \alpha_{\sigma(n)}\}; \{\dx_{\sigma(1)},\cdots
\dx_{\sigma(n)}\}
\right) y^{\sum_{i<j} \alpha_{\sigma(i)\sigma(j)}}\, , \nonumber \\
\een
where the sum runs over all permutations $\sigma$ of $\{1,2,\cdots n\}$.
The function $F$ is determined recursively by the equations
\be \label{eseed}
F(\{\ta_1\}; \{\cc_1=0\})=1\, ,
\ee
\ben
 \label{enewrecur}
&& F(\{\ta_1,\cdots \ta_n\};  \{\cc_1, \cdots \cc_n\}) \nonumber \\
&=& \Theta\left(- \ta_{n-1, n} \cc_n\right) (-1)^{\Theta(-\ta_{n-1,n})}
F(\{\ta_1,\cdots \ta_{n-1}\};  \{\cc_1, \cdots \cc_{n-2}, \cc_{n-1}+\cc_n\}) \nonumber \\
&& + \sum_{k=0}^{n-3} F(\{\ta_1, \cdots \ta_k, \ta_{k+1}+\cdots \ta_{n-1}+\lambda_k \ta_n\};
\{\cc_1, \cdots \cc_k, \cc_{k+1}+\cdots \cc_n\}) \qquad \qquad \qquad \nonumber \\
&& \times G(\ta_{k+1}, \cdots \ta_{n-1}, \lambda_k \ta_n) \, \Theta\left(-
\sum_{i,j\atop k+1\le i<j\le n-1} \ta_{ij} \sum_{i,j\atop k+1\le i<j\le n} \ta_{ij}\right)
{\rm sign}\left(\sum_{i=k+1}^{n-1} \ta_{kn}\right) \, ,
\een
where $\tilde \alpha_i$ are arbitrary elements of $\VV$, and 
\be
\lambda_k = - \sum_{i,j\atop k+1\le i<j\le n-1} \ta_{ij} / \sum_{i=k+1}^{n-1} \ta_{in}\, .
\ee
$\Theta(x)$ is the step function
\be \label{edefthx}
\Theta(x) = \begin{cases}0 \quad \hbox{for} \quad x<0\\
1  \quad \hbox{for} \quad x>0 \end{cases} \, ,
\ee
and the functions $G$ are determined from another set of recursion relations to be given below.
Once the functions $G$ are known, \refb{enewrecur} can be used to construct
$F$ recursively starting with the initial value  in \refb{eseed}.

Physically $F\left(\{\alpha_{\sigma(1)},\cdots \alpha_{\sigma(n)}\}; 
\{\dx_{\sigma(1)},\cdots
\dx_{\sigma(n)}\}
\right)$ represents the sum of the signs $s(P)$ introduced in \S\ref{sfirst} for all solutions $P$ in
which the $z_i$'s are ordered as $z_{\sigma(1)} < z_{\sigma(2)} < \cdots < z_{\sigma(n)}$.
$G\left(\{\alpha_{\sigma(1)},\cdots \alpha_{\sigma(n)}\}\right)$  has a similar %bp
interpretation except that in obtaining solutions to \refb{eor1} we set all the $c_i$'s to zero.
A useful property of $G$ %bp
 is that $G(\ha_1,\cdots \ha_n)$ for any
$\{\ha_i\}$ vanishes if the set $\{1,\cdots n\}$ can be divided into two sets $A$ and $B$ such that
$\ha_{ij}>0$ for all $i\in A$ and $j\in B$\cite{unpublished}.

Note that the step and sign functions appearing on the right hand side of 
\refb{enewrecur} and similar factors appearing in the recursion relations for $G$ to be
described below, will be ambiguous if their argument vanishes. The remedy given in
\cite{1302.5498} is again to work with a deformed set of parameters. This will be
reviewed in \S\ref{sdeform}.

We now turn to the definition of the function
$G$. First of all $G(\ha_1,\cdots \ha_n)$ is non-zero 
only if $\sum_{i<j}\ha_{ij}=0$. Moreover $G$
vanishes if $n<3$. For $n=3$,
\be \label{eg3pre}
G(\ha_1, \ha_2, \ha_3) = \Theta(\ha_{12} \ha_{23}) \,
(-1)^{\Theta(\ha_{23})+1}\,, \qquad \ha_i\in \VV.
\ee
Another important property of $G$ is that it depends on
its arguments $\ha_i$ only via the DSZ products $\ha_{ij}=\langle \ha_i, \ha_j\rangle$.
Hence in the recursion relations  below, it is sufficient to specify at every stage
the $\ha_{ij}$'s appearing in the argument of $G$; we do not need to explicitly specify the
$\ha_i$'s (nor even show the existence of $\ha_i$'s satisfying the specified 
$\ha_{ij}$'s).
For four
or more arguments, $G$ satisfies the recursion relation
\be \label{egrecur}
G(\ha_1,\cdots \ha_m) = (-1)^{1+\Theta(\ha_{m-1,m})} \Theta\left(-
\ha_{m-1,m} \sum_{i=1}^{m-1}\ha_{im}
\right) \, G(\ca_1, \cdots \ca_{m-1}) + \sum_B \Delta G_B\, ,
\ee
where $\ca_i$'s for $1\le i \le (m-1)$ satisfy
\ben \label{edefca}
&& \ca_{m-3, m-1} = -\ca_{m-1,m-3}= 
\ha_{m-3, m-1} + \sum_{i=1}^{m-1} \ha_{im}, \nonumber \\
&& \ca_{ij} = \ha_{ij} \quad \hbox{for} \quad (i,j) \ne (m-1, m-3) \, \hbox{or} \, 
(m-3, m-1)\, .
\een
The sum over $B$ in \refb{egrecur}
runs over all subsets of $\ha_1,\cdots \ha_m$ of  three or more
{\it consecutive elements} and containing either $\ha_m$, or both
$\ha_{m-1}$ and $\ha_{m-3}$ (or all three of them).
$\Delta G_B$ is given as follows.
First of all we introduce auxiliary elements $\ha_1',\cdots \ha_m'$ of $\VV$ satisfying
\ben \label{edefa}
&& \ha_{im}'= -\ha_{mi}'=\mu_B \, \ha_{im} \quad \hbox{for $i=1,2,\cdots m-1$}, 
\nonumber \\
&& \ha_{m-3,m-1}' = - \ha_{m-1,m-3}'= \ha_{m-3,m-1}
+ (1-\mu_B) \sum_{i=1}^{m-1} \ha_{im}\, , \nonumber \\
&& \ha'_{ij} =\ha_{ij} \quad \hbox{for all other $i,j$}\, ,
\een
where $\mu_B$ takes different values for different choices of $B$ and will be specified below.
\begin{enumerate}
\item For $B=\{\ha_{m-2},\ha_{m-1},\ha_m\}$ and $m>4$, we have
\be
\mu_B = -\ha_{m-2,m-1}  / (\ha_{m-1,m} + \ha_{m-2,m})\, ,
\ee
and
\ben
\label{escal1}
\Delta G_B &=& G(\ha_1', \cdots \ha'_{m-3}, \ha'_{m-2}+\ha'_{m-1}+\ha'_{m}) \times
G(\ha'_{m-2}, \ha'_{m-1}, \ha'_m) \,   \nonumber \\
&& \times \, {\rm sign} \left(\ha_{m-1,m} + \ha_{m-2,m}
\right) \, \Theta\left(- 
\left(\ha_{m-2,m-1} +\ha_{m-1,m} + \ha_{m-2,m}\right)
\ha_{m-2,m-1}
\right)\, . \nonumber \\
\een
\item For $B=\{\ha_k,\cdots \ha_m\}$ for $3\le k\le (m-3)$ we have
\be 
\mu_B = \left( \sum_{i,j\atop
k\le i<j\le m-1} \ha_{ij} + \sum_{i=1}^{m-1} \ha_{im} \right) \Big/ \sum_{i=1}^{k-1} \ha_{im} \, ,
\ee
and
\ben
\label{escal2}
\Delta G_B &=& G\left(\ha_1', \cdots \ha'_{k-1}, \sum_{i=k}^m \ha'_{i}\right) \times
G(\ha'_{k}, \ha'_{k+1}, \cdots, \ha'_m) \, {\rm sign} \left(-\sum_{i=1}^{k-1} \ha_{im}
\right)\nonumber \\
&& \times \Theta\left(-
\left(\sum_{i,j\atop k\le i<j\le m} \ha_{ij} \right)
\left( \sum_{i,j\atop k\le i<j\le m-1} \ha_{ij}
+  \sum_{i=1}^{m-1} \ha_{im} \right)
\right)\, .
\een
\item For $B=\{\ha_k,\cdots \ha_{m-1}\}$ for $2\le k\le (m-3)$ we have
\be 
\mu_B = \left( \sum_{i,j\atop k\le i<j\le m-1} \ha_{ij} + \sum_{i=1}^{m-1} \ha_{im} \right)
\Big/ \sum_{i=1}^{m-1} \ha_{im} \, ,
\ee
and
\ben
\label{escal3}
\Delta G_B &=& G\left(\ha_1', \cdots \ha'_{k-1}, \sum_{i=k}^{m-1} \ha'_{i}, \ha'_m\right) \times
G(\ha'_{k}, \ha'_{k+1}, \cdots, \ha'_{m-1}) \, {\rm sign} \left(-\sum_{i=1}^{m-1} \ha_{im}
\right) \nonumber \\ &&
\times \Theta\left(-
\left( \sum_{i,j\atop k\le i<j\le m-1} \ha_{ij}
+  \sum_{i=1}^{m-1} \ha_{im}\right)
\left( \sum_{i,j\atop k\le i<j\le m-1} \ha_{ij}
\right)
\right)
\, .
\een
\item For $B=\{\ha_2,\cdots \ha_m\}$ and $m>4$ we have
\be 
\mu_B = \left( \sum_{i,j\atop
2\le i<j\le m} \ha_{ij} +\ha_{1m} \right) \Big/  \ha_{1m} \, ,
\ee
and
\be \label{ejumpgs}  
\Delta G_B = \Theta\left(-
\left( \sum_{i,j\atop 2\le i<j\le m} \ha_{ij} \right)
\left( \sum_{i,j\atop 2\le i<j\le m} \ha_{ij}
+   \ha_{1m} \right)
\right) %\nonumber \\ && \times
{\rm sign}(-\ha_{1m}) \, G(\ha'_2, \cdots \ha'_m)\, .
\ee
\item For $m=4$ and $B=\{\ha_2,\ha_3, \ha_4\}$ we have
\be
\mu_B = - \ha_{23} / (\ha_{24}+\ha_{34})
\, , 
\ee
and
\be
\Delta G_B =
 \Theta\left(- \ha_{23}\left(\ha_{23} + \ha_{34}+\ha_{24}\right)\right)
\times {\rm sign}(\ha_{24} + \ha_{34}) \, G(\ha'_2, \ha'_3, \ha'_4)
\, . %\nonumber \\
\ee
\end{enumerate}
This finishes our description of $\Delta G_B$ in all cases. This in turn
defines $G$, and hence  $F$ and $g_C$.
The equivalence of the definitions of $\gref$ given in \S\ref{sfirst} and in this section
was proven in \cite{1302.5498}.

\subsection{Definition of the functions $H$}  \label{sdefh}

We now give the algorithm for constructing the functions 
$H(\{\beta_i\}; \{k_i\};y)$ appearing in \eqref{essp2}.
\begin{enumerate}
\item When the number of
$\beta_i$'s is less than three,  $H(\{\beta_i\}; \{k_i\};y)$ vanishes.
\item
For three or more number of $\beta_i$'s, 
observe that  the expression for $\QC(\sum_i k_i\beta_i; \zeta;y)$ given in
\refb{edefqbar},
\eqref{essp1} contains a term proportional to
$H(\{\beta_i\}; \{k_i\};y)\prod_i\OmS(\beta_i;t^{k_i})$
arising from the choice $m=1$ in  \eqref{edefqbar}, $n=1$,
$\alpha_1=\sum_i k_i\beta_i$, $m_1=1$
in \eqref{essp1}, and 
$m_i=k_i$ in the expression for
$\Omega_{\rm tot} (\sum_i k_i \beta_i; y,t)$ in eq.\eqref{essp2}.
We fix $H(\{\beta_i\}; \{k_i\};y)$ by demanding that the net
coefficient of the product $\prod_i \OmS(\beta_i;t^{k_i})$ in the
expression for $\QC(\sum_i k_i\beta_i; y,t)$ is a
Laurent polynomial in $y$, \i.e.\ a finite sum of the form
$\sum c_s y^s$ where $c_s$'s are constant and the sum over $s$ runs over
a finite set of positive and negative integers (including zero). 
This of course leaves open the possibility  of
adding to $H$ a
Laurent polynomial. This is resolved by using the {\it minimal
modification hypothesis}, which requires that $H$ must 
be symmetric under $y\to y^{-1}$ and vanish
as $y\to\infty$  \cite{1103.1887}.
We determine $H(\{\beta_i\}; \{m_i\};y)$ iteratively by
beginning with the $H$'s with three $\beta_i$'s and then 
determining successively the $H$'s with more $\beta_i$'s. 

For illustration, suppose that the coefficient of $\prod_i \OmS(\beta_i;t^{k_i})$
in some example was $(y^2+y^{-2}) / (y - y^{-1})^2$ before taking into account the
$H(\{\beta_i\}; \{k_i\};y)\prod_i\OmS(\beta_i;t^{k_i})$ term. Then 
we choose $H(\{\beta_i\}; \{k_i\};y)=-2/ (y-y^{-1})^2$ so that
 the total coefficient of $\prod_i \OmS(\beta_i;t^{k_i})$ becomes
$(y^2+y^{-2}-2) / (y - y^{-1})^2=1$, which is a Laurent polynomial in $y$.

\item
$H$ is expected 
to be independent of the
FI parameters and hence can be calculated for any
value of these parameters.
\end{enumerate}
A useful property of $H(\{\beta_1,\cdots \beta_n\};\{k_1,\cdots k_n\}; y)$
is that it vanishes if the set $\{1,\cdots n\}$ can be divided into two sets $A$ and $B$
such that $\langle \beta_i,\beta_j\rangle\ge 0$ for all $i\in A$ and $j\in B$\cite{unpublished}.

\subsection{Constraints on $\OmS(\alpha;t)$'s %as single centered indices
} \label{soms}

In the Coulomb branch formula $\OmS(\alpha;t)$ are unknown functions of $t$.
In this section we shall discuss some necessary conditions on $\alpha$ in order for
$\OmS(\alpha;t)$ {\it not} to vanish. It is important to remember, however, 
that these conditions should be used 
{\it after determining the functions $H(\{\beta_i\}; \{k_i\};y)$
following the procedure of \S\ref{sdefh} where we treat all the $\OmS(\alpha;t)$ for
$\gamma\in \Gamma^+$ to be non-vanishing and independent of each other.}

The first set of conditions on $\OmS(\alpha;t)$ are
\be
\label{OmSquivers}
\OmS(a\gamma_i+b \gamma_j;t)=\begin{cases}1 \quad \mbox{if $a=1, b=0$ or $a=0, b=1$}\\
0 \quad \mbox{otherwise}
\end{cases}
\ee
for any  linear combination of two  basis vectors $a\gamma_i+b\gamma_j$. 
$\OmS(\sum_i n_i\gamma_i;t)$ can be non-zero if at least 3 of the $n_i$'s are
non-zero, but only if the following condition holds. Let $\beta_1,\cdots \beta_n$
be the set of $n=\sum_i n_i$ vectors with $n_1$ of the $\beta_k$'s being $\gamma_1$,
$n_2$ of the $\beta_k$'s being $\gamma_2$ etc., and let 
$\beta_{k\ell} \equiv\langle \beta_k,\beta_\ell\rangle$. Then for $\OmS(\sum_i n_i\gamma_i;t)$ to
be non-zero there must exist $n$ vectors $\vec r_1,\cdots \vec r_n\in \mathbb{R}^3$
satisfying the equations
\be \label{ecoms}
\sum_{\ell=1}^n {\beta_{k\ell}\over |\vec r_k - \vec r_\ell|} = 0\, \quad \forall \quad k\, .
\ee
For $n=3$ this condition requires that the set of numbers $(\beta_{12}, \beta_{23}, \beta_{31})$
or $(-\beta_{12}, -\beta_{23}, -\beta_{31})$ are all positive and satisfy the triangle 
inequality\cite{0702146}. For $n\ge 4$ there is no such simple criterion but one can show that
\refb{ecoms} has no solutions if we can divide the set $\{1,\cdots n\}$ into two sets $A$ and $B$
such that $\beta_{k\ell}\ge 0$ for all $k\in A$ and $\ell\in B$\cite{unpublished}.
Further constraints
on $\OmS$ which follow from symmetry under quiver mutations
 will be discussed in \refb{econb4}. %bp

The Coulomb branch formula gives an expression for the Dolbeault polynomial
-- a function of two variables $y$ and $t$ 
-- in terms of $\OmS(\alpha;t)$'s which are functions of a single variable $t$. Furthermore
the Dolbeault polynomial depends on the FI parameters $\zeta$ but the $\OmS(\alpha;t)$'s
do not. So by knowing $\OmS(\alpha;t)$ for one $\zeta$ one can deduce  
$\QC(\gamma; \zeta;y,t)$ for any other values of the FI parameters. A practical way of computing 
$\OmS(\alpha;t)$ is to specialize  \refb{essp1}, to $y=1/t=v^{1/2}$, where
the left-hand side becomes proportional to the Hirzebruch $\chi_y$-genus $\chi(\cM,v)$ %jm
via eq.\refb{eHir}.
Inverting this relation allows to express   $\OmS(\alpha;t)$ in terms of the Hirzebruch $\chi_y$-genera 
$\chi(\cM(\gamma;\zeta), v)$ associated to subquivers, for a fixed choice of stability condition 
$\zeta$.\footnote{Note that the Coulomb branch formula
expresses $\QC(\gamma;\zeta;y,t)$ for $\gamma=\sum_i N_i \gamma_i$ in terms of
$\OmS(\alpha;t)$ for $\alpha=\sum_i p_i\gamma_i$,
 $0\le p_i\le N_i$. Thus to systematically calculate the $\OmS$'s from the knowledge of $\chi$
 we need to begin with the lowest rank quivers (minimum possible values of
 $\sum_i N_i$) and then work our way upwards. 
}

\subsection{Deformations of the input parameters} \label{sdeform}

As mentioned in \S\ref{sfirst} and \S\ref{second}, in order to determine $\gref$
unambiguously we need to work with a deformed set of parameters. In this
subsection we shall review this prescription.
However we would like to mention at the outset that while
the deformed parameters are to be used in eqs.\refb{eor1} and \refb{ecou0}
for enumerating the solutions and
computing $s(P)$, and also computing the step and sign functions appearing in
various equations in \S\ref{second},
the exponent of 
$y$ 
in \refb{eindex1} or \refb{eqforg}
is always computed with the undeformed $\alpha_{ij}$'s, related to 
the  undeformed $\gamma_{ij}$'s via \refb{esymp}.
\begin{enumerate}
\item In the first step we deform  
the original numbers $\gamma_{ij}$ by arbitrary small numbers:
\be \label{edeform1}
\gamma_{ij} \to \gamma_{ij} +\eps_1 \xi_{ij}\, ,
\ee
where $\eps_1$ is a small positive number and $\xi_{ij}$ are arbitrary but
sufficiently generic real numbers between
$-1$ and $1$ satisfying $\xi_{ij}=-\xi_{ji}$. In particular
for given $\eps_1$,
$\xi_{ij}$ should be such that any ordered set of vectors 
$\beta_1,\cdots \beta_s\in \Gamma^+$ with $s\ge 2$ and not all $\beta_k$'s parallel,
satisfying $\sum_k\beta_k=\sum_i n_i\gamma_i$
with $n_i\le N_i$, have the property that $\sum_{i<j} \beta_{ij}\ne 0$. 
This condition ensures that 
the set of equations  \refb{eor1} has no solutions where two or more centers coincide,
except if the $\alpha_i$'s associated with all the centers are parallel.
Furthermore $\eps_1$ should be sufficiently small so that for given $\xi_{ij}$
the above condition holds for any $\eps_1$ between the chosen value and 0 except possibly at
$\eps_1=0$. Since the undeformed $\beta_{ij}$'s are all integers, the last condition can be ensured
by taking $\eps_1$ to be less than $2 / N (N-1)$ where 
$N=\sum_i N_i$ is the rank of the quiver.

The deformation parameters $\eps_1$ and $\xi_{ij}$, once chosen, must be kept fixed  
throughout the calculation,
{\it e.g.} for $\gref$ with various arguments as well as for calculating the functions $H$
using the procedure described in \S\ref{sdefh}
we must use the same 
deformation parameters.

\item The above deformation does not completely remove the ambiguity in enumerating the
solutions to \refb{eor1} since two or more $z_i$'s whose associated $\alpha_i$'s 
are parallel can still
coincide. Furthermore if the original vector $\gamma$ is not primitive 
then for the arguments $c_i$
of $\gref$ in \refb{essp1}, $\sum_i' c_i$ may
 vanish where $\sum'_i$ denotes sum over a proper
 subset of the elements. This allows a solution where two sets of $z_i$'s have infinite
 separation from each other. To avoid such situations we carry out a second deformation 
 in which we choose some arbitrary ordering of the $\alpha_i$'s and
  deform the $\alpha_{ij}$'s and $c_i$'s in the argument of $\gref$ to
 \be \label{edeform2}
 \alpha_{ij} \to \alpha_{ij}+\eps_2 \eta_{ij}\, , \quad c_i\to c_i + \eps_2 f_i\, .
 \ee
Here $\eps_2$ is a small positive number,
and 
$f_i$ and $\eta_{ij}$'s are arbitrary but sufficiently generic
real numbers between
$-1$ and $1$ satisfying $\eta_{ij}=-\eta_{ji}>0$ for $j>i$ and $\sum_i f_i=0$. 
For given $\eps_1$, $\xi_{ij}$ and $\eps_2$,
$\eta_{ij}$ and $f_i$ should be such that
$\sum_{i,j\in A; i<j}\alpha_{\sigma(i)\sigma(j)}$ is non-zero for any permutation
$\sigma$ and any
subset $A$ of $1,\cdots n$ containing two or more elements, and
$\sum_{i\in B} c_i$ should be non-zero for any non-empty proper
subset $B$ of $1,\cdots n$. 
These conditions ensure that with the deformed parameters the set of equations
\refb{eor1} has no solutions where two or more centers coincide or where one or more
centers  get infinitely separated from the rest.
Furthermore $\eps_2$ should be sufficiently small so that for given $\eps_1$,
$\xi_{ij}$ and $\eta_{ij}$, the conditions mentioned above hold for all
$\eps_2$ between the chosen value and 0, except possibly at $\eps_2=0$.
If we choose the original values of $\zeta_i$'s and the deformations
$\eps_1\xi_{ij}$ to be integer multiples of $1/\Lambda$ for some large number
$\Lambda$, then the last condition can be ensured by taking $\eps_2$ to be
less that $2/(n (n-1) \Lambda)$.

Note that this set of deformations is `local' {\it e.g.} specific to the evaluation
of $\gref$ with a specific set of arguments. For evaluating $\gref$ with another set of
arguments we use another set of deformation parameters $\eps_2$ and $\eta_{ij}$
satisfying the conditions described above. However different terms contributing to
a given $\gref$ must be computed with the same set of deformation parameters.

\item This finishes the procedure that needs to be followed for the evaluation of
$\gref$ given in \S\ref{sfirst}. However this still leaves some ambiguity in the
analysis of \S\ref{second} since
even after we deform the parameters $\alpha_{ij}$ and $c_i$ using the two-step
process described above, one or more of the arguments of the
step and sign functions on the right hand side of \refb{enewrecur}, and similar functions
appearing in the recursion relations for $G$,
may vanish accidentally making these functions
ambiguous. These occur on codimension $\ge 1$ subspaces of the space of the deformation
parameters $\eps_2$ and $\alpha_{ij}$ and we can adjust these parameters to stay away
from these subspaces. 
\end{enumerate}
It was argued in \S3.2 of \cite{1302.5498} that $\QC$ computed from this procedure
is independent of the choice of the deformation parameters $\eps_1\xi_{ij}$ and
$\eps_2\eta_{ij}$.

For quivers without
oriented loops an alternative version of the Coulomb branch formula that does not require
the deformations given in \refb{edeform1},  \refb{edeform2} was given in %bp
\cite{1212.0410}. We do not know of an extension of this to the general case.

\subsection{Mathematica code}

The algorithm for computing the Dolbeault polynomial $\QC$ using the Coulomb branch formula
was implemented 
in a Mathematica package called {\tt CoulombHiggs.m}, whose first version was described in Appendix A of  \cite{1302.5498}. The latest release and updated instructions can be downloaded from 
\begin{center}
\href{http://www.lpthe.jussieu.fr/~pioline/computing.html}{{\tt http://www.lpthe.jussieu.fr/$\sim$pioline/computing.html}}
\end{center} 
The package contains many other commands to compute e.g. the Coulomb indices $g_C$ or the Reineke formula mentioned in \S\ref{sec_reineke} of this review. Here we describe the main
command for computing Coulomb branch formula.
After copying the file  {\tt CoulombHiggs.m}  in the  same directory as the notebook, evaluate

\medskip

$\mathematica{0.9}{SetDirectory[NotebookDirectory[]]; <<CoulombHiggs.m}{CoulombHiggs v2.1 - A 
package for evaluating quiver invariants using the Coulomb and Higgs branch formulae.}{1}$

\noindent The   main routine for evaluating $\QC$ is {\tt \color{functioncolor} CoulombBranchFormula}, 
whose usage is as follows:

\mathematica{0.9}{Simplify[CoulombBranchFormula[4\{\{0, 1, -1\},\{-1, 0, 
   1\}, \{1, -1, 0\}\}, \{1/2, 1/6, -2/3\}, \{1, 1, 1\}]]  
     }
   {$ 2+ \frac{1}{y^2}+y^2 +  \text{OmS}(\{1,1,1\},y,t) $
}{2}

\noindent This routine computes $\QC(\gamma;\zeta; y, t)$. 
The first argument corresponds to the matrix of DSZ products $\gamma_{ij}$ (an antisymmetric matrix of integers), the second to the FI parameters $\zeta_i$ (a vector of rational numbers) and
 the third to the dimension vector $N_i$ (a vector of integers). 
 The code allows for the possibility of $y$-dependence in $\OmS$
(called
OmS in the code), but
for generic superpotential the single-centered
indices $\OmS(\gamma;y,t)$ can be taken to be  
independent of $y$.  The constraint \refb{ecoms} has not been incorporated in the
code, so some of the $\OmS$'s appearing in the expression for $\QC$ may actually
be zero.

One of the bottlenecks in running  the
code is due to the need for the deformations described in \refb{edeform1} and \refb{edeform2}.
The current version uses randomly generated small rational numbers for these deformations. This 
works well for small number of nodes, but when the number of nodes becomes
large (say $\ge 6$) occasionally the arguments of one of the step functions 
vanish accidentally. In that case, the step function evaluates to $1/2$ and generates a warning message. In such cases it is advisable to run the code again:  since each run  uses different random  deformations, typically the warning messages will disappear. 

Irrespective of this issue, the code could certainly be optimized by compiling and/or parallelizing the computation of the Coulomb indices (the performance of the version 2.1 released along with this review is already greatly improved with respect to previous versions, by avoiding to evaluate Coulomb indices which end up being multiplied by zero).

\sectiono{Aspects of the Coulomb branch formula} \label{saspects}

In this section we shall discuss various aspects of the Coulomb branch formula.

\subsection{Quivers without oriented loops and Reineke's formula \label{sec_reineke}}

For quivers without oriented loops, the necessary condition \refb{ecoms} always fails and hence all 
$\OmS(\alpha)$'s vanish except for the $\OmS(\gamma_i)$'s associated to basis vectors $\gamma_i$, which are all equal to 1
according to \refb{OmSquivers}.  $\QC$ 
computed from \refb{essp1} is therefore $t$-independent, and the cohomology is supported in degree $(p,p)$. %bp
Furthermore by suitably choosing the deformation parameters described in \S\ref{sdeform}
one can simplify the formula for $\gref$ given in \S\ref{second}. The 
deformations are chosen as follows. First we choose the $\xi_{ij}$'s in
\refb{edeform1} such that the deformed
quiver also does not have oriented loop. In this case the $\alpha_i$'s appearing in the
argument of $\gref$ in \refb{essp1} 
have the property that they can be ordered such that $\alpha_{ij}\ge 0$
for $i<j$. We choose this ordering and then deform them as in \refb{edeform2} with
$\eta_{ij}>0$ for $i<j$ so that the deformed $\alpha_{ij}$'s are all positive for
$i<j$.
One finds that in this case the  functions  $G$ appearing in the right hand side of
\refb{enewrecur} all vanish due to the property mentioned in the
paragraph below \refb{edefthx} and the recursion relation for $F$ reduces to 
just the first term on the right hand
side of \refb{enewrecur}. This gives\footnote{If we had chosen a different set of 
deformation parameters then
the intermediate steps will be more complicated but the final results will be the same.}
\be \label{efnoloop}
F(\{\ta_1,\cdots \ta_n\};  \{\cc_1, \cdots \cc_n\}) 
=\prod_{k=2}^n \left\{
\Theta\left(- \ta_{k-1, k} (\cc_k+\cdots \cc_n)\right) (-1)^{\Theta(-\ta_{k-1,k})}
\right\}\, .
\ee
Furthermore all the relevant $H$-functions also vanish due to the property mentioned at
the end of \S\ref{sdefh}.
It was shown in \S4 of \cite{1302.5498} that the Coulomb branch formula
computed with this expression for $F$
reduces to Reineke's formula\cite{1043.17010} 
for the Poincar\'e polynomial of the moduli space of quivers without
oriented loops.

\subsection{Wall-crossing} 
It was shown in \S4.1 and \S5.4 of \cite{1103.1887} that the Coulomb branch formula satisfies
the wall-crossing formula given in \cite{1011.1258} (and further elaborated in
\cite{1103.0261,1107.0723,1110.4847}), provided the single-centered invariants
$\OmS(\gamma)$ stay constant across the wall.
 
In the context of quivers the wall-crossing formula can be stated as follows. 
First note that \refb{enzeta}
gives $\zeta\cdot\gamma=0$ for $\gamma=\sum_i N_i\gamma_i$.
A quiver carrying charge vector $\gamma$ 
hits a wall of marginal stability if  the FI
parameters take values such that 
$\zeta\cdot \gamma'=0$ for another vector 
$\gamma'\not\parallel \gamma$. 
If we denote by $\wt\Gamma^+$ the intersection of $\Gamma^+$ with the
plane spanned by $\gamma$ and $\gamma'$ then it follows  
that for any vector $\alpha\in\wt \Gamma^+$, 
$\zeta\cdot\alpha$ vanishes on the wall. 
Let $\zeta^L$, $\zeta^R$ denote two different points in the FI parameter space on two sides of the
marginal stability wall, satisfying the conditions \refb{enzeta} and:
\be \label{eside1}
\langle \beta, \gamma\rangle \, \, (\zeta^L\cdot\beta) > 0, 
\quad 
 \langle \beta, \gamma\rangle \, \, (\zeta^R\cdot\beta) < 0, \quad \forall \, \, \beta\in \wt\Gamma^+\, .
  \ee
For 
any vector $\alpha\in \wt\Gamma^+$ let us define
\be
\zeta_\alpha \equiv \zeta - (\langle (\gamma - \alpha) , \gamma_1 \rangle, 
\cdots \langle (\gamma - \alpha) , \gamma_K\rangle) ( \langle \gamma, \alpha\rangle )^{-1} 
\zeta \cdot \alpha\, .
\ee
Note that $\zeta_\alpha\cdot \alpha=0$ automatically. 
For $\zeta$ close to the wall of marginal stability,  $\zeta_\alpha$ is close to $\zeta$
since $\zeta\cdot\alpha\simeq 0$. Furthermore $\zeta_\alpha$ is on the `same side'
of the  wall as $\zeta$ in the sense that \refb{eside1} holds with
$\gamma$ replaced by $\alpha$ and $\zeta$ replaced by $\zeta_\alpha$.
With this definition of $\zeta_\alpha$ 
the wall-crossing formula of \cite{1011.1258} can be stated as\footnote{The definition of
$\zeta_\alpha$ was not given explicitly before, but it follows from the analysis if
\S4.1 of \cite{1103.1887} in a straightforward manner.}
\be \label{ewall}
\bQC(\gamma,\zeta^L;y,t) = \sum_n {1\over n!} \sum_{\alpha_i\in \wt\Gamma^+\atop \sum_i \alpha_i 
= \gamma}  \gref(\alpha_1,\cdots \alpha_n; c_1^R, \cdots c_n^R; y) \prod_{i=1}^n
\bQC(\alpha_i,\zeta_{\alpha_i}^R;y,t)\, ,
\ee
where 
\be
c_i^R =  \zeta^R \cdot \alpha_i\, . 
\ee
The sum over  $\{\alpha_i\}$ in \refb{ewall}
runs over all
{\it ordered} set of vectors $\alpha_1,\cdots \alpha_n\in\wt\Gamma^+$ satisfying 
$\sum_i \alpha_i=\gamma$. 
Furthermore, even if the original quiver $\gamma$ has oriented loops, 
the $\alpha_i$'s appearing in the argument of $\gref$ on the right hand side of
\refb{ewall} can be partially ordered so that $\alpha_{ij}\ge 0$ for $i<j$. Hence by
deforming them so that $\alpha_{ij}>0$ for $i<j$,
we can use \refb{eqforg} with
the expression for $F$ given in
\refb{efnoloop} for computing $\gref$.
The equivalence of this wall-crossing formula  
and the Kontsevich-Soibelman wall-crossing formula\cite{Kontsevich:2008}
was proven in
\cite{1112.2515,1212.0410}.

\subsection{Generalized quiver mutations}

The Coulomb branch formula \refb{essp1} is conjectured to be invariant under a general
mutation symmetry\cite{1309.7053}. To describe this conjecture in its most general form 
we need a slight  generalization of this formula as follows:
\begin{enumerate}
\item We take the $\OmS$'s to depend on $y$ and denote them by $\OmS(\alpha;y,t)$. In all the
formul\ae\ in \S\ref{smain} we replace $\OmS(\alpha;t^m)$ by $\OmS(\alpha; y^m, t^m)$. %bp
\item We relax the constraints \refb{OmSquivers} and \refb{ecoms}. 
\end{enumerate}
We now pick a particular node $k$ and define $\Omega_{n,s}$ and $M$ via
\be
\label{eq:M}
\OmS(\ell\gamma_k;y,t) = \sum_{n,s } \Omega_{n,s}(\ell\gamma_k) y^n t^s, \qquad
M \equiv \sum_{\ell\geq 1}
 \sum_{n,s} \ell^2 \, \Omega_{n,s}(\ell \gamma_k) \ .
\ee
We further define
\ben
\label{mutDSZgen0}
%\begin{split}
\gamma'_i&=&\begin{cases}
-\gamma_k \quad \hbox{if $i=k$} \\
\gamma_i + M \, {\rm max}(0,\varepsilon \gamma_{ik})\, \gamma_k \quad  \hbox{if $i\neq k$}
\end{cases} \nonumber
\\
\gamma'_{ij} &=& 
\begin{cases}
-\gamma_{ij} \quad  \mbox{if}\quad i=k \quad \mbox{or}\quad  j=k \\
 \ \gamma_{ij} + M\, {\rm max}(0, \gamma_{ik} \gamma_{kj})\, {\rm sign}(\gamma_{kj}) \quad 
 \mbox{if} \quad i,j\neq k
 \end{cases}\nonumber
 \\
\zeta'_i&=& \begin{cases}
-\zeta_k \quad \hbox{if $i=k$} \\
\zeta_i+M\, {\rm max}(0, \varepsilon \gamma_{ik}) \, \zeta_k \quad \hbox{for  $i\ne k$},
 \end{cases}\nonumber
\\
N'_i&=&\begin{cases}
-N_k+ M \sum_{j\neq k} N_j \, {\rm max}(0,\varepsilon \gamma_{jk}) \quad \hbox{if $i=k$} \\
N_i \quad  \hbox{if $i\neq k$}
\end{cases}
%\end{split}
\een
where $\varepsilon$ takes values $\pm 1$ with the + sign corresponding to right mutation and
$-$ sign corresponding to left mutation. These transformation laws, which will be called the
generalized mutation transformations, guarantee that
\be
\gamma\equiv \sum_i N_i \gamma_i = \sum_i N_i'\gamma_i'\, .
\ee
We also define $\OmS'(\alpha;y,t)$ through the relations
\be
\label{egench00}
\OmS(\alpha;y,t) =
\begin{cases}
 \OmS'\left(\alpha+ M\, {\rm max}(0, \varepsilon \langle \alpha,\gamma_k\rangle)\, \gamma_k
; y,t\right) \quad \hbox{for $\alpha \not\parallel \gamma_k$} \\
\OmS'\left( - \alpha; y, t\right) \quad \hbox{for $\alpha\parallel \gamma_k$}
\end{cases}\, .
\ee
In \cite{1309.7053}  we conjectured
that $\QC(\gamma;\zeta;y,t)$ produced by the Coulomb branch formula is invariant
under the generalized mutation transformation: 
\be
\label{mutinv0}
\QC(\gamma; \zeta; y; t) =
\begin{cases}\QC'(\gamma; \zeta'; y; t)  \quad \hbox{if $\gamma\not\parallel \gamma_k$}\\
\QC'(-\gamma;\zeta;y,t)  \quad \hbox{if $\gamma\parallel \gamma_k$}\,,
\end{cases}
\ee
provided the following conditions hold:
\be
\label{einequal}
\varepsilon\, \zeta_k<0\, ,
\ee
\be \label{eposcon0}
\Omega_{n,s}(\ell\gamma_k)\geq 0\ \forall\  \ell>0\ ,\quad
\Omega_{n,s}(\ell\gamma_k) = 0 \, \,  {\rm for}
\, \, \ell > \ell_{\rm Max}\, ,
\ee
for some $\ell_{\rm Max}$.
In \eqref{mutinv0}, it is understood that in computing the l.h.s. we have to express
$\gamma$ as $\sum_iN_i\gamma_i$  treating $\gamma_i$'s as the basis vectors 
and apply
the Coulomb branch formula \eqref{essp1}, \eqref{essp2} 
with  single centered indices $\OmS(\alpha;y,t)$,
while in 
computing the r.h.s. we have to express
$\gamma$ as $\sum_i N'_i\gamma'_i$ treating $\gamma'_i$'s as the basis vectors and then
apply the Coulomb branch formula \eqref{essp1}, \eqref{essp2}
with  single centered indices $\OmS'(\alpha;y,t)$.
Since $\OmS'(\alpha';y,t)=0$ unless $\alpha'=\sum_i n_i''\gamma_i'$ with $n_i''\ge 0$,
the transformation law \refb{egench00} forces the correponding $\OmS(\alpha;y,t)$'s also to vanish.
This can be made explicit by expressing \refb{egench00} as
\be \label{econb1}
\OmS\left(\sum_i n_i \gamma_i\right) = \OmS\left(\sum_i n_i'' \gamma_i'\right)\, ,
\ee
where
\ben \label{econb2}
n_i'' &=& 
n_i \quad \hbox{for} \quad i\ne k\, , \nonumber \\
n_k'' &=& -n_k + M \, \sum_{j\neq k} n_j \, {\rm max}(0,\varepsilon \gamma_{jk}) 
- M \, {\rm max}\left(0,\varepsilon \sum_{j\neq k} \,  n_j \, \gamma_{jk}\right) \, .
\een
Thus the condition $n_k''\ge 0$ for non-vanishing $\OmS'$ translates to the following condition
for getting non-vanishing $\OmS(\sum_i n_i\gamma_i)$:
\be \label{econb3}
n_k \le M \, \sum_{j\neq k} n_j \, {\rm max}(0,\varepsilon \gamma_{jk}) 
- M \, {\rm max}\left(0,\varepsilon \sum_{j\neq k} \,  n_j \, \gamma_{jk}\right) \, .
\ee

For $M=1$ these transformation rules reduce to the usual mutation tranformation
rules discussed in \cite{zbMATH05573998,Kontsevich:2008,zbMATH05848698,0309191}. 
If we further restrict to quivers with generic superpotentials where
$\OmS$'s become $y$ independent, then 
\refb{egench00} shows that $\OmS'$'s are also $y$-independent. 
This is consistent with the fact
that under mutation, a generic superpotential transforms into a generic superpotential.
Note that the fact that the Coulomb branch formula with $y$-independent 
$\OmS$'s transforms under mutation to
the Coulomb branch formula with $y$-independent $\OmS'$'s is a non-trivial test of the Coulomb branch formula itself.  %bp
Furthermore since both left and right mutations are 
known to be symmetries of the quiver, the condition \refb{econb3} %bp
now gives a constraint on the $\OmS$'s associated with the physical quivers. We can in fact
make these constraints stronger by requiring $\OmS'$ to satisfy \refb{OmSquivers}.
This gives the following condition on $\OmS(\alpha)$:
\ben \label{econb4}
\OmS\left(\sum_i n_i\gamma_i\right) &=& 0 \, \, \hbox{unless} \, \,  
n_k \le  \, \sum_{j\neq k} n_j \, {\rm max}(0,\varepsilon \gamma_{jk}) 
- \, {\rm max}\left(0,\varepsilon \sum_{j\neq k} \,  n_j \, \gamma_{jk}\right)  \, \, \forall \, \, k;
\vareps=\pm 1\, , \nonumber \\
\OmS(a\gamma_i + b \gamma_j + c\gamma_k) &=& 0 \, \, 
\hbox{unless} \, \,  
c <  \, a\, {\rm max}(0,\varepsilon \gamma_{ik}) + \, b\, {\rm max}(0,\varepsilon \gamma_{jk}) 
-  \, {\rm max}\left(0,\varepsilon (a \gamma_{ik} + b \gamma_{jk})\right) \nonumber \\
&& \qquad \qquad \, \, \forall \, \, \hbox{distinct} \, \, i,j,k;
 \, a,b>0\, ;  \, \, \vareps=\pm 1\, . %\nonumber \\
\een

\subsection{Other tests}

The Coulomb branch formula passes several other
consistency tests. 
\begin{enumerate}
\item It was shown in \cite{1207.0821,1207.2230} that the Coulomb branch formula 
correctly reproduces the cohomology
of cyclic abelian quivers of the form
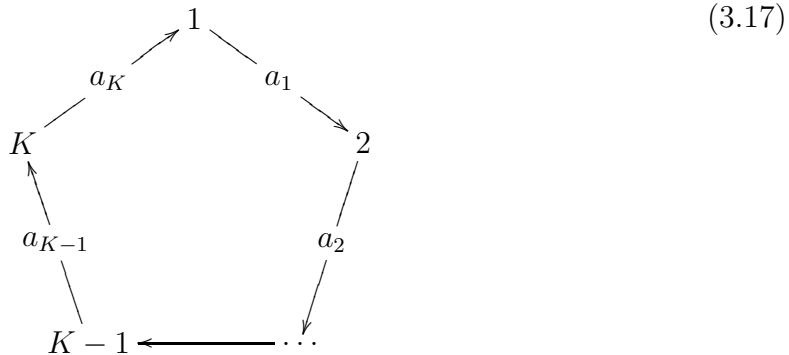
\be
\begin{xy} 0;<1pt,0pt>:<0pt,-1pt>:: 
(65,0) *+{1} ="0",
(129,47) *+{2} ="1",
(105,123) *+{\dots} ="2",
(25,123) *+{K-1} ="3",
(0,47) *+{K} ="4",
"0", {\ar|*+{a_1}"1"},
"1", {\ar|*+{a_2}"2"},
"2", {\ar"3"},
"3", {\ar|*+{a_{K-1}}"4"},
"4", {\ar|*+{a_K}"0"},
\end{xy}\\
\ee
Let us consider FI parameters in the range
\be
\zeta_1,\cdots \zeta_{K-1}>0, \qquad \zeta_K<0\, .
\ee
The result for the Dolbeault polynomial of this quiver, computed using the
Coulomb branch formula, takes the form
\be \label{ecycle2bbss}
\begin{split}
Q_C(\gamma_1+&\cdots +\gamma_K;y,t) \simeq \Omega_{\rm S}
(\gamma_1+\cdots +\gamma_K;t)
\\&
+(-1)^{K-1 + \sum_{\ell} a_\ell} (y-y^{-1})^{-K+1} \,
\sum_{\sigma_1=\pm 1, \sigma_2=\pm 1,
\cdots \sigma_K=\pm 1\atop 
\sign\left[{\sum_{\ell=1}^{K}} a_\ell\sigma_\ell\right] =
- \sigma_K}\,  \left(\prod_{\ell=1}^{K-1} \sigma_\ell\right)\, 
y^{\sum_{\ell=1}^K \sigma_\ell a_\ell} \ ,
\end{split}
\ee
where $\simeq$ denotes that this formula computes correctly the 
coefficients of the non-positive
powers of $y$. The coefficient of the positive powers of $y$ can be found using
the $y\to 1/y$ symmetry. This agrees with the result 
obtained directly from the analysis of the quiver moduli space %bp
in \cite{1205.5023,1205.6511}.

\item The Coulomb branch formula was tested in many other explicit examples in
\S5 and \S6 of \cite{1207.2230}. For example, consider the quiver
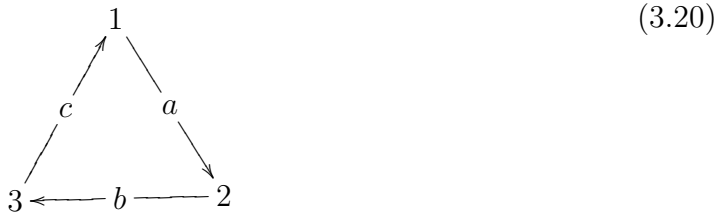
\be
\label{xy3node}
\begin{xy} 0;<1pt,0pt>:<0pt,-1pt>:: 
(38,0) *+{1} ="0",
(79,67) *+{2} ="1",
(0,69) *+{3} ="2",
"0", {\ar|*+{a}"1"},
"2", {\ar|*+{c}"0"},
"1", {\ar|*+{b}"2"},
\end{xy}
\ee
with 
rank (1,1,2), FI parameters
\be 
\label{ena1}
\quad \zeta_1>0, \quad \zeta_1+\zeta_2>0, \quad \zeta_2<0, 
\quad \zeta_3\to 0^-\, ,
\ee
and  positive integers
$a,b,c$ satisfying 
\be \label{erange}
 a<2c, \quad b< c\, , \quad k\equiv a+2b-2c>0\ .
\ee
The Coulomb branch formula gives the result
\be \label{ena6}
\begin{split}
Q_C&(\gamma_1+\gamma_2+2\gamma_3;y,t)\\
=& \Omega_{\rm S}(\gamma_1+\gamma_2+2\gamma_3;t)
+ (y - y^{-1})^{-3}\, (y+y^{-1})^{-1}\, \,
\bigg\{ y^{-k+1} -  y^{k-1} \\
& \qquad \qquad  + {1\over 4} (k-1) (y + y^{-1})^2 (y-y^{-1})
+{1\over 4} (-1)^{k/2} (y - y^{-1})^3 \bigg\}
\quad \hbox{for $k$ even}
\\
=&\Omega_{\rm S}(\gamma_1+\gamma_2+2\gamma_3;t)
+ (y - y^{-1})^{-3}\, (y+y^{-1})^{-1}\, \bigg\{ y^{k-1} - y^{-k+1}
\\
& \qquad \qquad - {1\over 2}(k-1) (y^2 - y^{-2})\bigg\}  
\quad \hbox{for $k$ odd}\, .
\end{split}
\ee
This agrees with the result of explicit computation of the Dolbeault polynomial of the quiver moduli space. %bp

\item For quivers with oriented loops the Reineke formula given in \cite{1043.17010},
called the Higgs branch formula in \cite{1302.5498}, no longer computes the %bp
 Poincar\'e polynomial of the quiver moduli space. Instead, it computes the cohomology of the (non-compact) space obtained by imposing only the D-term constraints
given by the first set of equations in \refb{emodi1}, together with the effect of taking the quotient by the
$\prod_i U(N_i)$ gauge groups. 
Let us call this the embedding space.
Under suitable conditions the F-term constraints, given by the
second set of equations in \refb{emodi1}, can be regarded as  sections %bp
of line bundles with positive curvature. Using repeated application of
Lefschetz hyperplane theorem we can then relate the
Hodge  numbers $h^{p,q}$ of the quiver moduli space 
$\cM$ to the Hodge number of the 
embedding space for low values of $(p+q)$.
Since the former is conjecturally 
given by the Coulomb branch formula whereas the latter is given by the
Higgs branch formula, this gives a specific relation between the Coulomb and the Higgs branch
formul\ae. The precise relation takes the following form\cite{unpublished}. If we take the
expression for $\QC(\gamma;\zeta; y,t)$ then the coefficients of the powers of $y^{-1}$,
that do not have any explicit dependence on unknown $\OmS$'s after using \refb{OmSquivers},
must match with the corresponding coefficients in the Higgs branch formula. It is however important
that {\it we do not use \refb{ecoms} to set some of the $\OmS$'s to zero for making this comparison,
although
we are allowed to set $\OmS(\sum_i n_i\gamma_i;t)$ to zero if the subquiver 
given by the dimension vector $(n_1,\cdots n_K)$ does not contain a closed loop}.
The technical reason for this is that even if a subquiver fails to satisfy \refb{ecoms} but has
oriented loop, we can construct a gauge invariant superpotential involving the variables
$\phi_{\ell k;\alpha;ss'}$ associated with the subquiver.
This correspondence between the Coulomb and the Higgs branch formula has been tested in
a number of examples\cite{unpublished}.

\end{enumerate}

\subsection{Variants of the conjecture}

The version of the conjecture stated in \S\ref{s2} is in its strongest form. There are
however some more conservative versions of this conjecture which we describe below.

\begin{enumerate}
\item If we set $t=1$ then the Dolbeault polynomial \refb{eDol}
reduces to the Poincar\'e polynomial
containing information about the Betti numbers only. The Coulomb branch formula 
\refb{essp1} for $t=1$ now expresses the Betti numbers in terms of 
$\OmS(\gamma; t=1)$.\footnote{Note that this has to be done after we have determined
the functions $H$ following the procedure described in \S\ref{sdefh}.}
The latter in turn can be related to the Euler number $\chi(\cM)$
by setting $y=t=1$ on both sides of \refb{essp1}, since in that case \refb{eDol} gives 
$Q(\cM; y=1, t=1) = (-1)^d \chi(\cM)$. Thus the Coulomb branch formula specialized to $t=1$ can be
viewed as an expression for the Betti numbers of $\cM$ in terms of the Euler numbers
of the moduli space of $\cQ(\gamma;\zeta)$ and its subquivers. This weak form
of the conjecture is the one which is directly related to and motivated by
black hole quantum mechanics.  %bp

\item Another version of the conjecture allows the
$\OmS$'s to depend both on $t$ and $y$.
In this case, in order to ensure consistency with the wall-crossing formula, 
all factors of $\OmS(\alpha; t^m)$ have to be replaced
by $\OmS(\alpha; y^m, t^m)$ in \eqref{essp2}. For fixed values of 
the FI parameters $\zeta$, the $\OmS(\alpha; y,t)'s$ are just a repackaging
of the Dolbeault polynomials $Q(\alpha;\zeta;y,t)$. However, unlike the latter,
the former are independent of the FI parameters. 
We can in fact regard the Coulomb branch formula with $(t,y)$-dependent $\OmS$'s as the most general solution to the wall-crossing formula. This formula may be valid more generally, {\it e.g.} for quivers with non-generic 
superpotentials. %bp

\end{enumerate}

\subsection{Application to framed BPS indices}  

Beyond its utility for computing the cohomology of quiver moduli spaces, and applications to black hole counting, the Coulomb branch formula can also be used to compute framed BPS indices associated to line defects in $\cN=2$ supersymmetric field theories \cite{Gaiotto:2010be}. 
For a wide class of models, the framed BPS spectrum can be obtained by adjoining an additional `framing node' $\gamma_{F}$ to the original quiver that governs the BPS spectrum in the bulk. 
The prescription of \cite{1308.6829} is that the framed BPS indices are computed using
the Coulomb branch formula, with single-centered invariants $\OmS(\alpha)$ equal to
0 whenever the charge vector $\alpha$ has non-zero support on $\gamma_F$, except for $\OmS(\gamma_F)$ itself, which is set equal to 1. It was checked in various examples that the resulting generating functions for framed invariants satisfy the Operator Product Expansion algebra for line defects \cite{1308.6829} and agree for $y=1$  with the generating functions 
computed in \cite{Gaiotto:2010be}. It is important to stress that the framed BPS indices obtained by this prescription differ from the BPS indices describing the cohomology of the corresponding framed quiver with generic superpotential, as the latter requires a different set of single centered invariants $\OmS(\alpha)$, which in general do not vanish when $\alpha$ has
support on the framing node.

\section*{Acknowledgments}
We are grateful to Clay Cordova, Kentaro Hori, Andy Neitzke and Piljin Yi for valuable discussions. B.P. is grateful to HRI for hospitality during the final stage of this work. 
The work of A.S.  was
supported in part by the project   12-R\&D-HRI-5.02-0303
and the J. C. Bose fellowship of 
the Department of Science and Technology, India.

\end{document}